\documentclass[twocolumn,showpacs,preprintnumbers,amsmath,amssymb,superscriptaddress]{revtex4}
\usepackage{graphicx}
\usepackage{dcolumn}
\usepackage{bm}
\usepackage{soul}
\usepackage{color}
\usepackage{epstopdf}
\usepackage[version=3]{mhchem}
\usepackage{lipsum}
\usepackage[outercaption]{sidecap}
\usepackage{floatrow}

\begin{document}

\title{The pH-dependent electrostatic interaction of a metal nanoparticle with the MS2 virus-like particles}
\author{Anh D. Phan}
\affiliation{Faculty of Materials Science and Engineering, Phenikaa Institute for Advanced Study, Phenikaa University, Hanoi 100000, Vietnam}
\affiliation{Faculty of Information Technology, Artificial Intelligence Laboratory, Phenikaa University, Hanoi 100000, Vietnam}
\email{anh.phanduc@phenikaa-uni.edu.vn}
\author{Trinh X. Hoang}
\affiliation{Institute of Physics, Vietnam Academy of Science and Technology, 10 Dao Tan, Ba Dinh, Hanoi, Vietnam}%
\date{\today}

\begin{abstract}
The electrostatic interaction of metal nanoparticles with viruses is attracting great interest due to their antiviral activity and their role in enhancing the detection of viruses at ultra-low concentrations. We model the MS2 virus devoid of its single strand RNA core using a core-shell model. The dependence of the inner and outer surface charge density on the pH is taken into account in our model of the interaction. Varying the pH causes a change in the sign  of the outer surface charge leading to the attractive-repulsive transition in the electrostatic interaction between the MS2 virus and metal nanoparticle at pH = 4.
\end{abstract}
\maketitle
\section{Introduction}
A significant increase in morbidity and mortality attributed to viral pathogens has sustained a worldwide pursuit in the investigation of viruses. For instance, the World Health Organization ranked lower respiratory infections to be one of the top 10 leading causes of death, taking over 3 million lives in 2016\cite{6}.  Although vaccines and/or antiviral drugs have been developed to inhibit viral replication or propagation for various viral species over the past century, genetic mutations inevitably occur which may result in treatment resistant strains of pathogenic viruses.  Some viruses, called bacteriophages, infect bacterial cells instead of human cells and can safely be used to treat bacterial infections. Bacteriophages, such as the MS2, PRD1 and $\phi$X174 viruses, have been widely exploited to study properties \cite{30,31,32,33,34,35} and behaviors of viruses at ultra-low concentrations, under which they are hard to be detected. Scientists in Ref. \cite{30,31,32,33,34,35} have investigated (i) dependence of virus or/of virus-like particle aggregation on pH and salinity, (ii) applicability of standard  Derjaguin-Landau-Verwey-Overbeek (DLVO) theory for viruses, and (iii) various peculiar features of electrokinetics of viruses. A thorough understanding of bacteriophages is important for proposing appropriate medical treatments and for environmental applications.   

Recent applications of metal nanostructures in biology and medicine have demonstrated the potential for nanotechnology to revolutionize these fields. For example, the presence of gold nanoshells significantly amplifies the signal in the micro-spherical whispering gallery mode of the MS2 virus and shifts the resonance microcavity wavelength \cite{7}. Gold nanoparticles have also been used to improve the detection of the hepatitis B virus and to quantify its concentration \cite{8}. Virus detection is possible by high-resolution imaging performed by the sharp metal probe of a scanning tunneling microscope (STM) \cite{11}. Recently, scientists have employed silver nanoparticles to inhibit the HIV-1 virus from binding to its host cells \cite{9}, and to inhibit hepatitis B virus replication \cite{10}. Howver, when metal nanoparticles having a virucidal effect conjugate with bacteriophages for antimicrobial activity, the selected bacteriophages could be inactive antibacterial agents. Their mutual interaction may be fundamental for the development of future applications.

A theoretical approach for investigating soft particles, including bacteria and viruses, based on the Poisson-Boltzmann equations was introduced by Ohshima for the first time in 1994 \cite{12}. The results of this approach are qualitatively and quantitatively in good agreement with experiments \cite{13,14,15}. When soft particles are immersed in electrolyte solutions, the theory gives us predictions of their electrostatics and electrokinetics. Using the predicted electrostatics, one can estimate electrostatic interactions between two particles \cite{24,18}. 
Duval \emph{et. al} \cite{30} developed a generic theory for the evaluation of electrostatic interactions between soft particles that possibly consist of successive layers defined by distinct electrostatic, structural and protolytic properties. This theory overcomes some of the limitations of the original formulations by Ohshima, and it does not suffer from the Deryagin and Debye-Hückel approximations, classically assumed in literature and inapplicable for moderate to highly charged particles who size is comparable to Debye layer thickness. The approximate Ohshima's model has been reported to provide qualitative interpretation of experimental data \cite{14,17}.
In 2013, Phan \emph{et. al} \cite{13} proposed a core-shell structure for a soft particle associated with some extensions of the Ohshima theory to describe the bacteriophage MS2 virus. This extended the earlier work of  Nguyen and her coworkers \cite{16}. In Phan's model, soft particles consist of a charged hard core having a fixed volume charge density surrounded by a charged outer layer. Despite of the fact that the calculations agree qualitatively well with experimental results \cite{16} and Phan's virus model was subsequently refined over several years \cite{19,20,21,22,23}, the MS2 virus model ignores the  inner and outer surface charges.

In this paper, we investigate the electrostatic interaction of a metal nanoparticle with the MS2 virus. Our virus has no RNA core but effects of surface charges on the electrostatic force are considered. Since the surface charge densities were experimentally proved to be strongly dependent on solution pH \cite{5}, we determine variations of the electrostatic interactions with pH.

\section{Theoretical background}
\begin{figure}[htp]
\includegraphics[width=8.5cm]{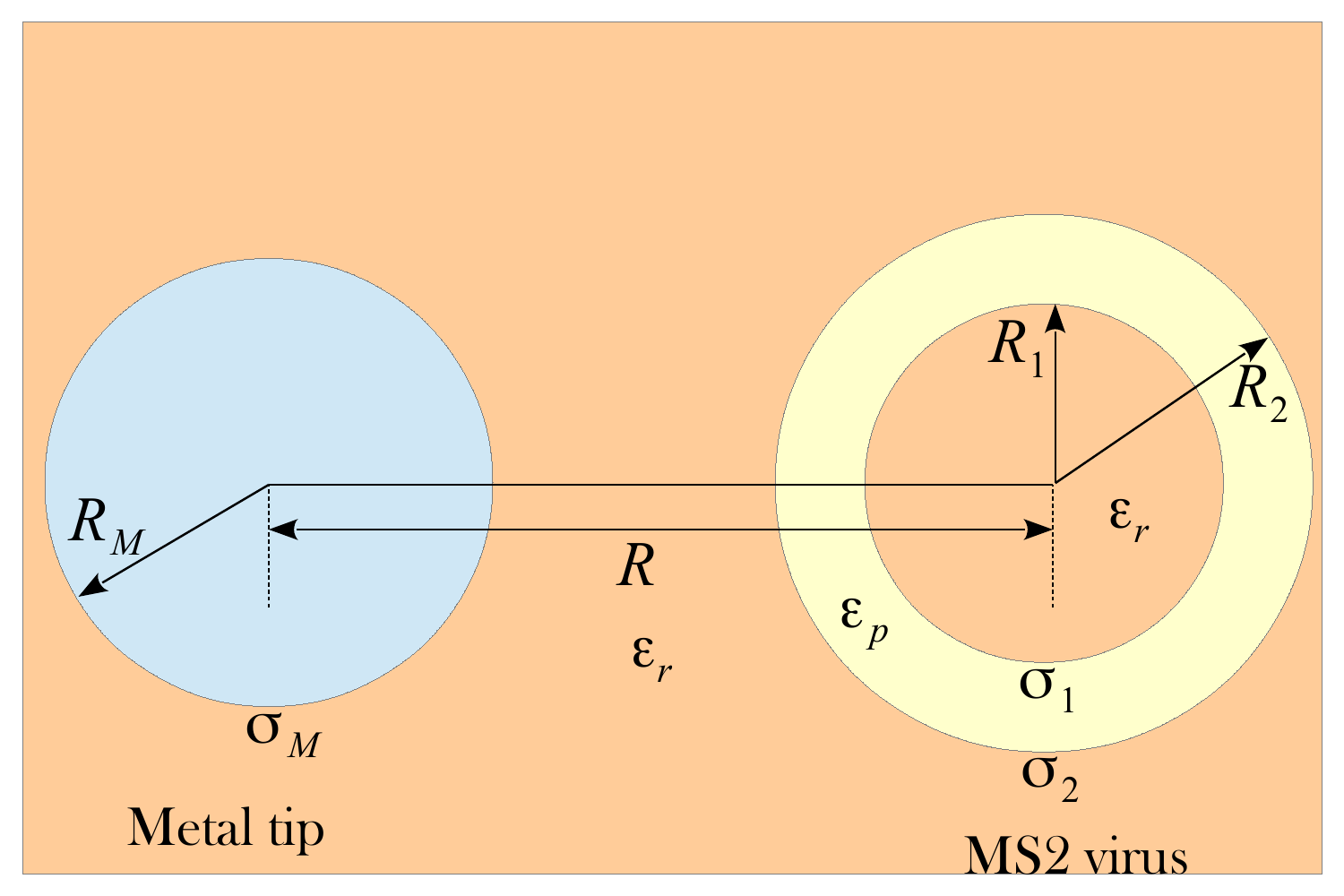}
\caption{\label{fig:1}(Color online) Illustration of the electrostatic model for the interaction of a spherical metal tip and a MS2 virus, both immersed in an electrolyte solution with permittivity $\varepsilon_r$.}
\end{figure}

In this section we describe and analyze our model of the electrostatic interaction between a metal tip and a MS2 virus immersed in an electrolyte solution as depicted in Fig. \ref{fig:1}. The virus are modelled by a core-shell model. In most viruses, the core of  a virus contains DNA or RNA and is encapsulated by a viral protein layer, called the capsid. In our work, we consider the MS2 virus without a RNA core in order to match the conditions and use the computed charges from a prior work \cite{5}. The electric interaction is strongly dependent on the surface potentials on the two objects. Knowing the electrostatic potential when these species are isolated from each other is a key to approximating the force of the interaction.

\subsection{Electrostatic potential profile of an isolated metal tip}
We consider a spherical tip possessing a radius $R_M = 15$ nm and a surface charge density $\sigma_M = -0.0025$ ($C/m^2$) \cite{14}. The electric potential distribution of the metal tip obeys the linearized Poisson-Boltzmann (PB) equations

\begin{eqnarray}
\left\{ \begin{array}{rcl}
\cfrac{d^2\psi_M}{dr^2}+\cfrac{2}{r}\cfrac{d\psi_M}{dr} = \kappa^2\psi_M, & \
& R_M \le r < \infty \\
\cfrac{d^2\psi_M}{dr^2}+\cfrac{2}{r}\cfrac{d\psi_M}{dr} = 0, &  & 0 \le r < R_M \\ 
-\cfrac{\partial\psi_M}{\partial r}|_{r=R_M} = \cfrac{\sigma_M}{\varepsilon_0\varepsilon_r},
\end{array}\right.
\label{eq:1}
\end{eqnarray}
where $r$ is the radial distance to the center of the tip, $\varepsilon_r$ is the permittivity of the aqueous solution, $\varepsilon_0$ is the vacuum permittivity, $\kappa^2 = 2z^2e^2n/\varepsilon_r\varepsilon_0k_BT$ is the Debye-H\"uckel parameter \cite{1}, $k_B$ is the Boltzmann constant, $T$ is an ambient temperature, and $z$ and $n$ are the ionic valence and the ion concentration in solution, respectively. Solving Eq.(\ref{eq:1}) gives
\begin{eqnarray}
\psi_M(r) = \frac{\sigma_MR_M^2}{(\kappa R_M+1)\varepsilon_0\varepsilon_r}\frac{e^{-\kappa (r-R_M)}}{r}.
\label{eq:2}
\end{eqnarray}

Recall that our calculations are valid within the Debye-H\"uckel approximation which assumes a low potential. When the condition of this approximation is violated, the PB equation becomes non-linear, then one can use the approach proposed by Duval et. al \cite{30} to yield better quantitative predictions. Duval's approach is applicable to soft particles having any size and captures strong effects of  particle surface curvature on electric field distribution and on electrostatic interaction energy.

\subsection{Electrostatic potential profile of an isolated MS2 virus}
The MS2 virus is modeled as a core-shell soft sphere, which has an inner radius $R_1 = 10.5$ nm and an outer radius $R_2 = 14$ nm \cite{13,16}. The soft particle is devoid of its single strand RNA core. We assume that the protein shell contains channels allowing free ion exchange between the interior and exterior of the capsid, while free charges are prohibited in the capsid region. The assumptions have been used in many prior literatures \cite{2,4,25}. The linearized Poisson-Boltzmann (PB) equations and boundary conditions for the MS2 virus are \cite{2,4}

\begin{eqnarray}
\left\{ \begin{array}{rcl}
\cfrac{d^2\psi}{dr^2}+\cfrac{2}{r}\cfrac{d\psi}{dr} = \kappa^2\psi, & \
& R_2 \le r < \infty \\
\cfrac{d^2\psi}{dr^2}+\cfrac{2}{r}\cfrac{d\psi}{dr} = 0, &  & R_1 \le r < R_2 
\\ \cfrac{d^2\psi}{dr^2}+\cfrac{2}{r}\cfrac{d\psi}{dr} = \kappa^2\psi, &  & 0 \le r < R_1 \\
\varepsilon_0\varepsilon_r\cfrac{d\psi}{dr}|_{R=R_1^-}-\varepsilon_0\varepsilon_p\cfrac{d\psi}{dr}|_{R=R_1^+}=\sigma_1, \\
\varepsilon_0\varepsilon_p\cfrac{d\psi}{dr}|_{R=R_2^-}-\varepsilon_0\varepsilon_r\cfrac{d\psi}{dr}|_{R=R_2^+}=\sigma_2, 
\end{array}\right.
\label{eq:3}
\end{eqnarray}
where $r$ is the radial distance to the center of the virus, $\sigma_1$ and $\sigma_2$ are the surface charge density of inner and outer surfaces, respectively, and $\varepsilon_p\approx 5$ is the dielectric constant of the capsid. In the calculations, dielectric permittivities in the inner and outer components of the virus-like particle are supposed to be indential. 

The general solutions for the above PB equations are
\begin{eqnarray}
\left\{ \begin{array}{rcl}
\psi(r) = B\cfrac{e^{-\kappa r}}{r}, & \
& R_2 \le r < \infty \\
\psi(r) = \cfrac{C}{r} + D , &  & R_1 \le r < R_2 
\\ \psi(r) = A\cfrac{\sinh\left[-\kappa r\right]}{r}, &  & 0 \le r < R_1
\end{array}\right.
\label{eq:4}
\end{eqnarray}
The coefficients $A$, $B$, $C$, and $D$ in Eq.(\ref{eq:4}) are found by satisfying the two boundary conditions in Eq.(\ref{eq:3}). 

The temperature- and concentration-dependent permittivity of the solvent is given by \cite{3}

\begin{eqnarray}
\varepsilon_r &\equiv& \varepsilon_r(c,T) = \varepsilon_W(T)h(c),\nonumber\\
\varepsilon_W(T) &=& 249.4 - 0.788T + 7.2\times 10^{-4}T^2,\\
h(c) &=& 1 - 0.255c + 5.15\times 10^{-2}c^2-6.89\times 10^{-3}c^3,\nonumber
\label{13}
\end{eqnarray}
where $c$ is the ionic strength of the solution. The experiments on the MS2 virus of Ref.\cite{5}  were carried out at $T = 293$ $K$. We use the same conditions in our calculations.

\subsection{Interactions of the metal sphere with the MS2 virus}
Approximating to the leading order, the electrostatic free energy between the metal sphere and the virus has been intensively studied and can be expressed via the unperturbed surface potential \cite{24},
\begin{eqnarray}
V(R)&\approx&4\pi\varepsilon_0\varepsilon_rR_MR_2\psi_M(R_M)\psi(R_2)\frac{e^{-\kappa(R-R_2-R_M)}}{R}.\nonumber\\
\label{eq:5}
\end{eqnarray}
where $R$ is their center-to-center distance. From Eq.(\ref{eq:5}), one can calculate the force of the interaction, 
\begin{eqnarray}
F(R)=-\frac{\partial V(R)}{\partial R}.
\label{eq:6}
\end{eqnarray}

\section{Numerical results and discussion}

\begin{figure}[htp]
\includegraphics[width=8.5cm]{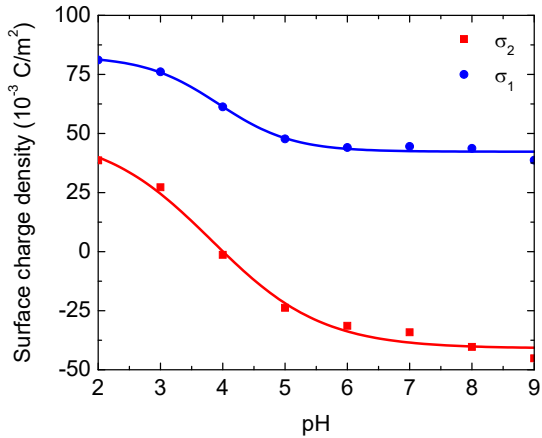}
\caption{\label{fig:2}(Color online) The pH-dependent surface charge density of the MS2 virus in absence of the RNA core. Data points derived from the analysis in Ref. \cite{5}. Lines correspond to the fit functions. }
\end{figure}

Figure \ref{fig:2} shows the dependence of the surface charge density for the inner and outer surfaces on pH. The data is obtained from analyzing Protein Data Bank (PDB) of the virus capsids in Ref.\cite{5}. Armanious and his coworkers \cite{5} reported the calculated charges of whole capsids and outer capsid surfaces as a function of pH. For simplification purpose, we simply subtract these two charge densities to find the inner surface charge density. The spatial dependence of the structural charges of the capsid is supposed to be the step-like function. Note that in a prior work \cite{31}, 
Langlet and coworkers developed a theoretical description for the spatial distribution of local volume charge density on particle radial position with adopting a tanh-like function (derived from cryoEM data). Clearly, our assumption is quite consistent with Langlet's calculations but our derivation remains less realistic as it ignores the spatial dependence of the density of ionogenic groups that are carried by the proteinaceous backbone of the virus-like particles.
The pH dependence of $\sigma_1$ and $sigma_2$ can be fitted using
\begin{eqnarray}
A_1+\frac{A_2-A_1}{1+10^{(\ce{logx0}-pH)p}},
\nonumber
\label{eq:7}
\end{eqnarray}
where $A_1$, $A_2$, $\ce{logx0}$, and $p$ are fit parameters. For $\sigma_1$, $A_1=42.31\times 10^{-3}$ $C/m^2$, $A_2 = 82.791\times 10^{-3}$ $C/m^2$, $\ce{logx0} = 3.93$, and $p = -0.73225$. For $\sigma_2$, $A_1=-40.94\times 10^{-3}$ $C/m^2$, $A_2 = 50.2226\times 10^{-3}$ $C/m^2$, $\ce{logx0} = 3.825$, and $p = -0.491$. While the outer surface charge density of the virus capsid changes from a positive value to a negative value as increasing pH, the inner surface charge density is positive over the considered range of pH. Remarkably, above a pH of 5 the surface charge density for the inner surface is essentially insensitive to pH. The pH-dependent variations of $\sigma_1$ and $\sigma_2$ are qualitatively, not quantitatively, similar to prior study \cite{25}. In Ref.\cite{25}, Podgornik and his coworkers also used PDB inputs to study the effects of pH on surface charge densities of the virus capsid. Additionally, several previous experiments \cite{15,16} proved that the presence of the RNA core in the MS2 virus hardly modifies the effective net surface charge. These findings suggest that in the pH range from 5 to 8 the outer surface charge density is fully responsible for the variations of the surface electrostatic potential of viruses, whether they have an RNA core or none.

\begin{figure}[htp]
\includegraphics[width=8.5cm]{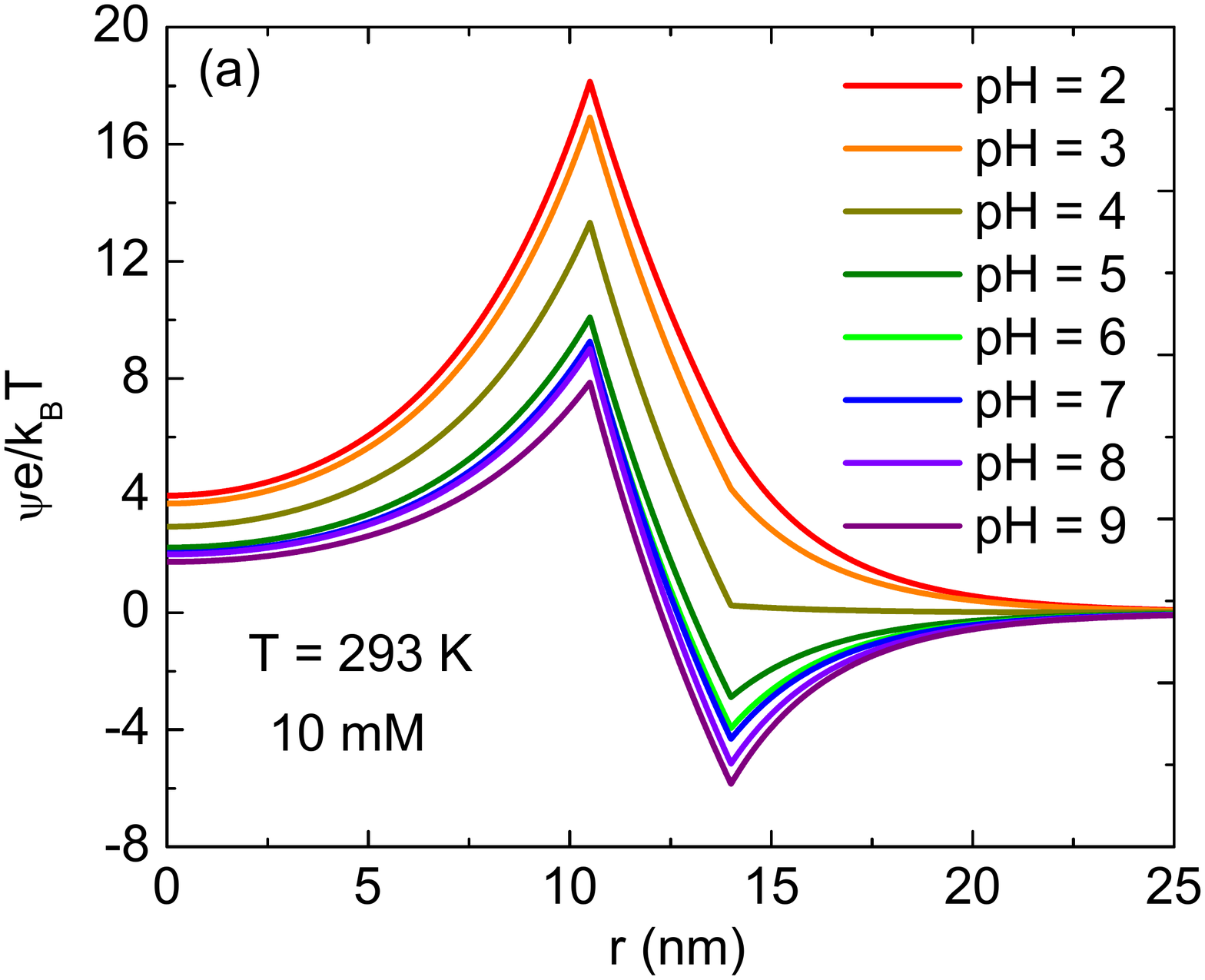}
\includegraphics[width=8.5cm]{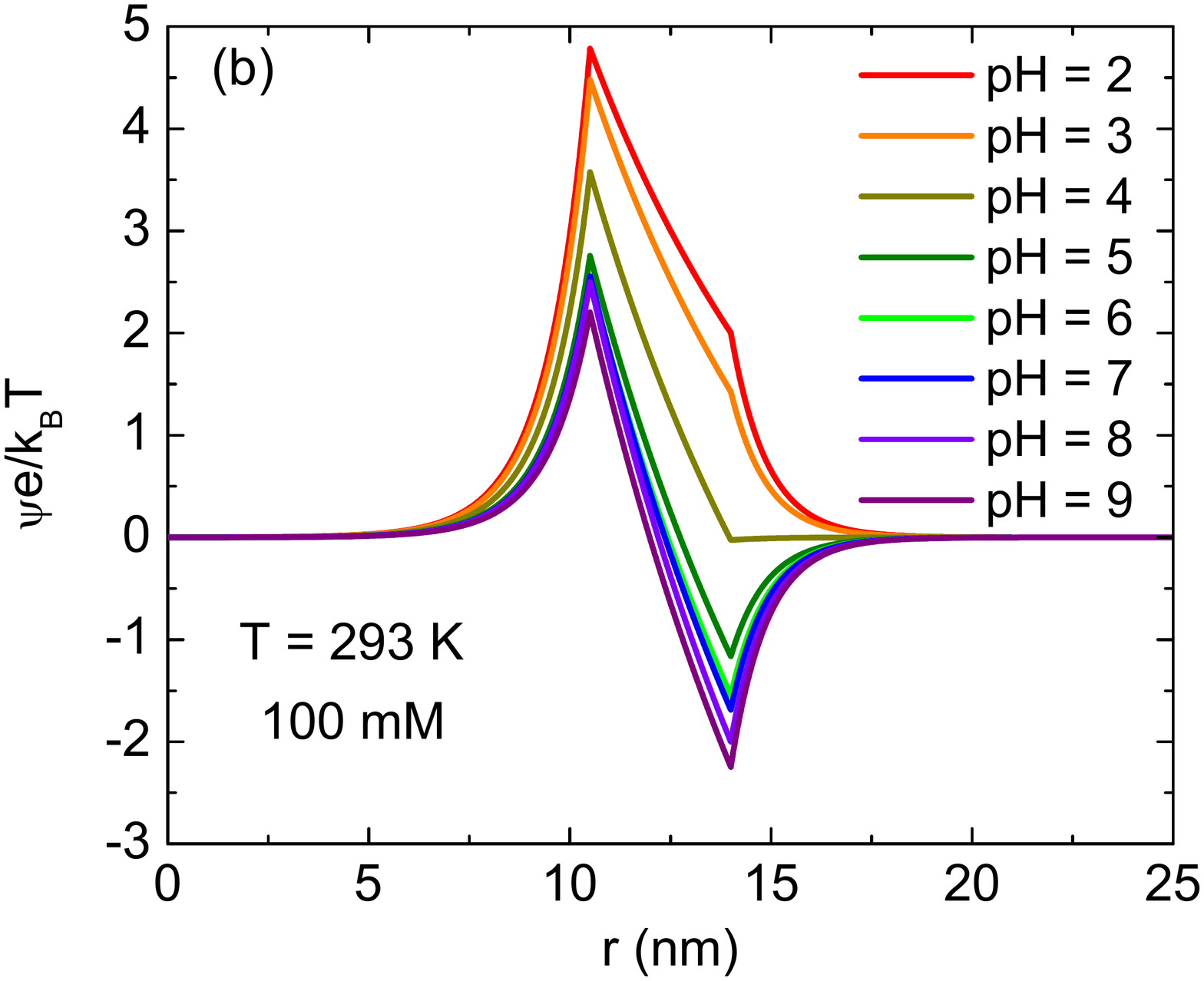}
\caption{\label{fig:3}(Color online) The pH-dependent electrostatic potential profile of the MS2 virus for a salt concentration of (a) $c= 10$ mM and (b) $c= 100$ mM.}
\end{figure}

The electrostatic potentials as a function of distance from the virus center at several pH values at the salt concentration of 10 and 100 mM are shown in Fig. \ref{fig:3}. Since the positive charge is localized on the inner surface of the virus capsid, the electrostatic potential is positive inside the virus. However, at low pH ($\leq 4$), the positive charge on the outer surface leads to the positive potential profile outside the virus. When the pH increases ($> 4$), the outer surface charge becomes negative, and thus the electrostatic potential reverses to negative values. Although the calculated values of $e\Psi(r)/k_BT$ in Fig. \ref{fig:3} are relatively large and thus the accuracy of the theoretical prediction is quantitatively reduced, it can be exploited to qualitatively explain experimental observations. In many previous studies \cite{32,33,34,35}, MS2-virus like particles begin to aggregate at at pH $\leq 4$. The finding suggest the electrostatic forces experience a replusive-to-attractive transition at pH = 4. At a given pH, a salt concentration increase leads to a decrease of $\varepsilon_r$ and a decay length, $1/\kappa$, of the electrostatic potentials. Consequently, both the exponential decay range and the magnitude of the potential at $c = 100$ mM are significantly reduced in comparison with those at $c = 10$ mM. This is a reason why we see the size of MS2-like virus particles aggregation at $c = 100$ mM much less than that at $c = 10$ mM \cite{33,34}. Authors in Ref. \cite{33} revealed that the stability of MS2-like virus particles against aggregation at $c = 10$ mM is strongly dependent  on purification process in experiments.

\begin{figure}[htp]
\includegraphics[width=8.5cm]{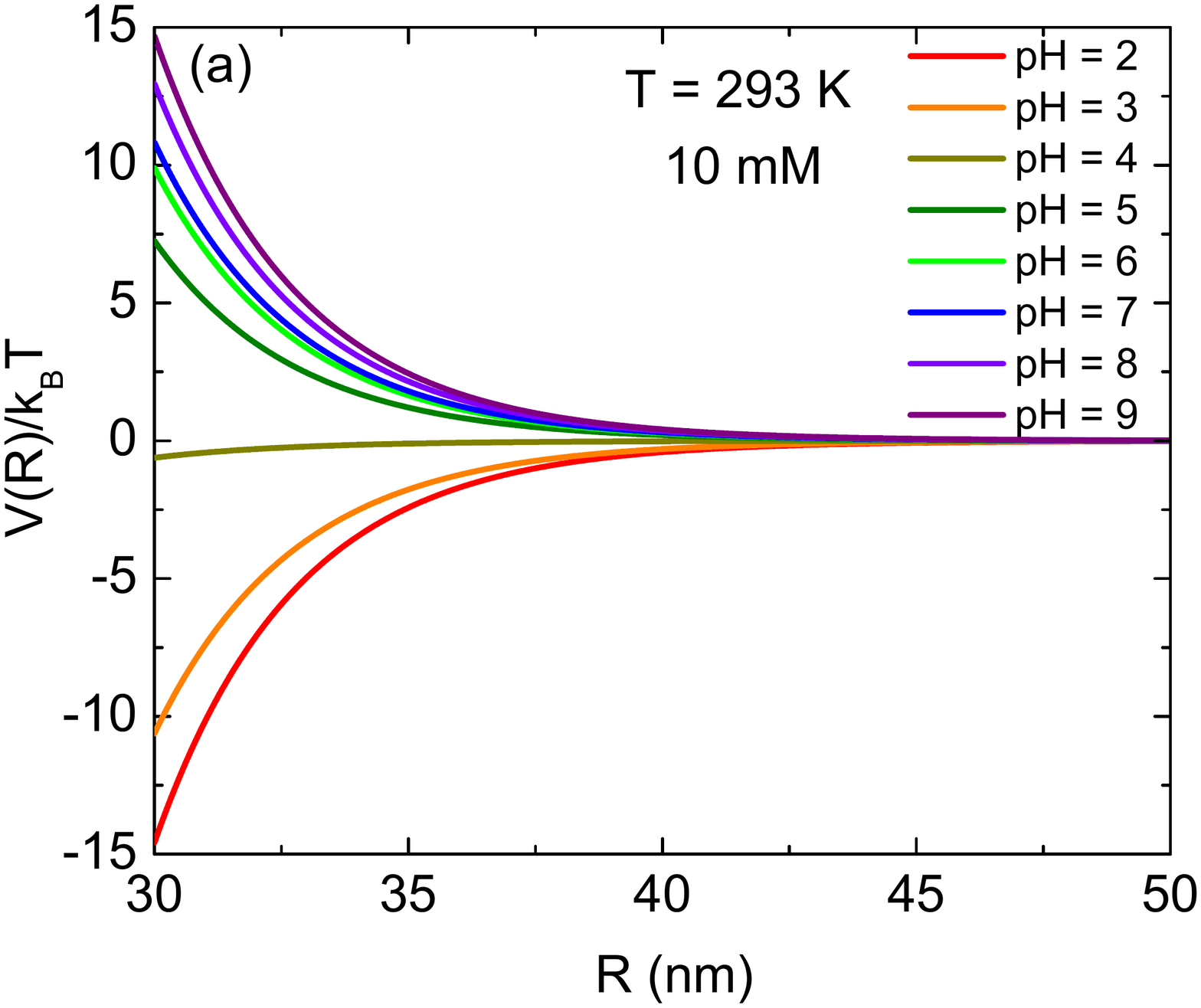}
\includegraphics[width=8.5cm]{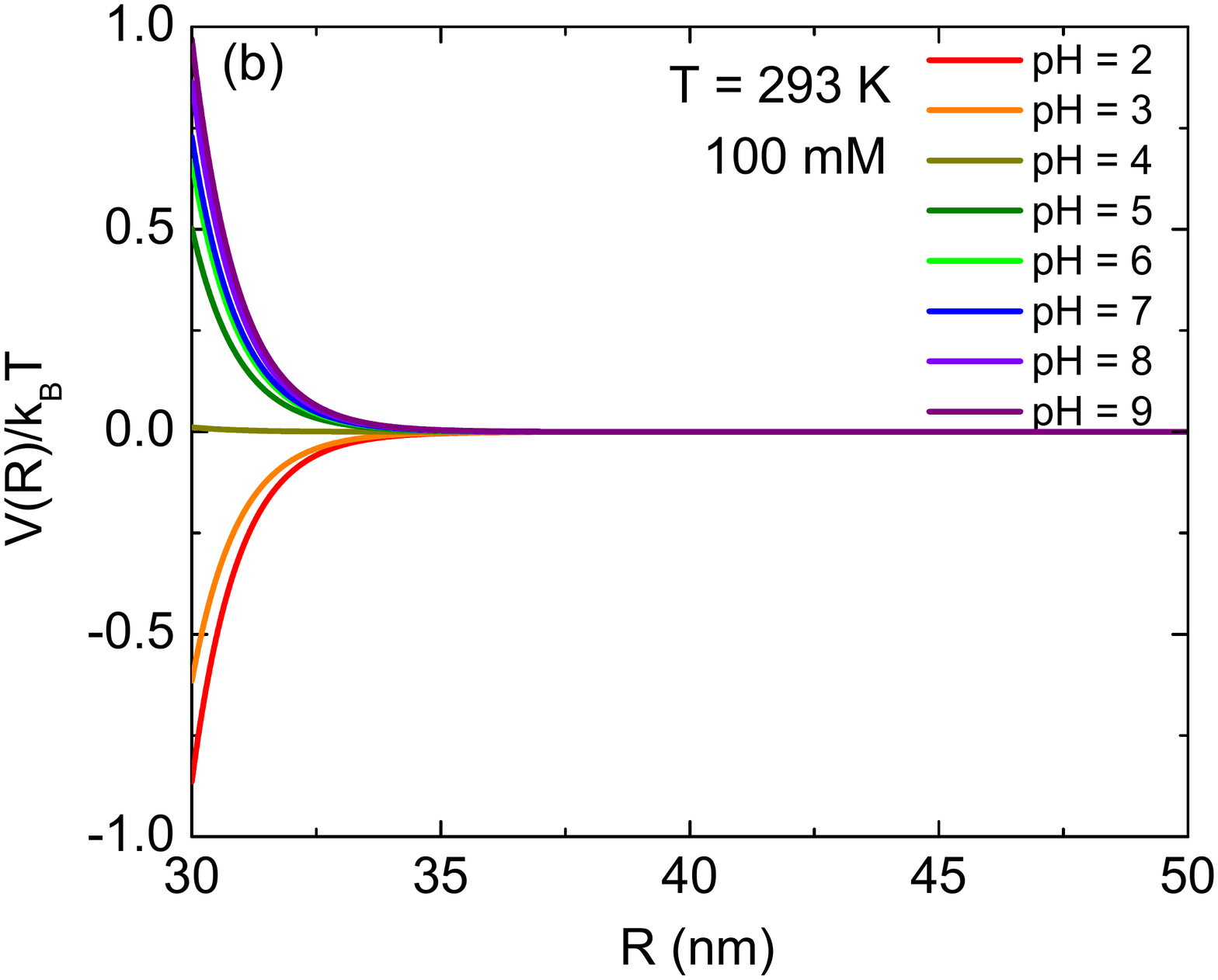}
\caption{\label{fig:4}(Color online) Electrostatic energy for the interaction of the MS2 virus with the metal tip as a function of center-center distances for several values of pH for a salt concentration of (a) $c= 10$ mM and (b) $c= 100$ mM. }
\end{figure}

Figure \ref{fig:4} shows the electrostatic interaction of the MS2 virus with the metal tip at various pH values as a function of the separation distance $R$ between two centers at the salt concentration of 10 and 100 mM. As can be expected from the change in the sign of the outer surface charge of the virus, a transition occurs in the electrostatic interaction from repulsive to attractive. One can approximately determine the radial distribution function between viruses and a metal sphere using $g(r)=e^{-V(r)/k_BT}$ \cite{26}. From this, the contact number can be also estimated. Near pH = 4, the outer surface of the virus capsid is neutral and the interaction becomes hard-sphere interaction. This behavior can be exploited in various interesting applications. First, adsorption or desorption of viruses from the metal nanoparticle is a remarkable signal for detecting a change in the pH. However, the metal nanoparticle in this case is very inert to viruses. Second,  laser irradiation can destroy adsorbed viruses via plasmonics-induced local heating or photothermal heating. Third, since silver nanoparticles can naturally inhibit viral replication and destroy bacteria, they have a potential to be used in therapeutic treatments that could selectively target infections in a particular location of the digestive tract. pH in human digestive tract ranges from 4.0 to 8.5. This is the region of pH studied in our work. For the targeted delivery application, the nanoparticles have to be bioconjugated with compatible biomolecules.

One can clearly observe substantial effects of salt concentration on the electrostatic interaction in Fig. \ref{fig:4}  when considering the same pH. In Fig. \ref{fig:4}a, for $c = 10$ mM, the electrostatic energy decays significantly to zero at $R \approx 45$ $nm$. While for larger salt concentration in the solution, the recovery range is about $R \approx 34$ $nm$ (much shorter than the dilute counterpart) as seen in Fig. \ref{fig:4}b. Additionally, the energy amplitude at $c = 100$ mM is approximately reduced by a factor of 15 compared to that at $c = 10$ mM. Note that the electrostatic potential profile of the MS2 virus decreases by a factor of 4 as increasing salt concentration from 10 to 100 mM. One can expect the variation of salt concentration also weakens the surface potential of the metal sphere.

\section{Conclusions}
We have presented our theoretical calculations for the pH-dependent electrostatic interactions between a metal tip and the MS2 virus. The MS2 virus without the RNA core is described using a core-shell model. The pH dependence of the inner and outer surface charge densities of the virus capsid has been discussed and compared to prior works. The outer surface charge was found to be fully responsible for the variations in the electrostatic interaction of the MS2 virus in absence of RNA core with the metal nanoparticle. Interestingly, the charge of the outer surface changes sign monotonically from positive to negative values at pH = 4. This leads to the repulsive-attractive transition seen at this pH.
\begin{acknowledgments}
This research is funded by Vietnam National Foundation for Science and
Technology Development (NAFOSTED) under Grant No. 103.01-2016.61. The authors would like to thank Dr. Dominik Domin for a careful reading and suggestions to this manuscript.

Conflict of Interest: The authors declare that they have no conflict of interest.
\end{acknowledgments}


\newpage 


\begin{thebibliography}{5}
\bibitem{6} https://www.who.int/news-room/fact-sheets/detail/the-top-10-causes-of-death

\bibitem{30} J. F. L. Duval, J. Merlin, and P. A. L. Narayana, Phys. Chem. Chem. Phys. {\bf 13}, 1037-1053 (2011).
\bibitem{31} J. Langlet, F. Gaboriaud, C. Gantzer, and J. F. L. Duval, Biophys. J. {\bf 94}, 3293-3312 (2008).
\bibitem{32} C. Dika, M. H. Ly-Chatain, G. Francius, J. F. L. Duval, and C. Gantzer, Colloids Surf. A Physicochem. Eng. Asp. {\bf 435}, 178-187 (2013).
\bibitem{33} C. Dika, C. Gantzer, A. Perrin, and J. F. L. Duval, Phys. Chem. Chem. Phys. {\bf 15}, 5691-5700 (2013).
\bibitem{34} C. Dika, J. F. L. Duval, H. M. Ly-Chatain, C. Merlin, and C. Gantzer, Appl. Environ. Microbiol. {\bf 77}, 4939-4948 (2011).

\bibitem{35} C. Dika, J.F.L. Duval, G. Francius, A. Perrin, C. Gantzer, J. Colloid Interface Sci. {\bf 446}, 327-334 (2015).
\bibitem{7} V. R. Dantham, S. Holler, V. Kolchenko, Z. Wan, and S. Arnold, Appl. Phys. Lett. {\bf 101}, 043704 (2012).
\bibitem{8} M. Shevtsov, L. Zhao , U. Protzer, and M. A. A. van de Klundert, Viruses {\bf 9}, 193 (2017).
\bibitem{11} J.-H.Lee, B.-K. Oh, and J.-W. Choi, Sensors {\bf 15}, 9915-9927 (2015).
\bibitem{9} J. L. Elechiguerra, J. L. Burt, J. R. Morones, A. Camacho-Bragado, X. Gao, H. H. Lara, and M. Jose Yacaman, J. Nanobiotechnology {\bf 3}, 6 (2005).
\bibitem{10} L. Lu, R. W.-Y. Sun, R. Chen, C.-K. Hui, C.-M. Ho, J. M. Luk, G. KK. Lau, and C.-M. Che, Antivir. Ther. {\bf 13}, 253-262 (2008).

\bibitem{12} H. Ohshima, J. Colloid Interface Sci. {\bf 163}, 474 (1994).
\bibitem{13} A. D. Phan, D. A. Tracy, T. L. Hoai Nguyen, N. A. Viet, T.-L. Phan, and T. H. Nguyen, J. Chem. Phys. {\bf 139}, 244908 (2013).
\bibitem{14} M. H.-Pérez, A. X. Cartagena-Rivera, A. L. Bozic, P. J. P. Carrillo, C. S. Martín, M. G. Mateu, A. Raman, R. Podgornik, and P. J. de Pablo, Nanoscale {\bf 7}, 17289 (2015).
\bibitem{15} V. I. Syngouna and C. V. Chrysikopoulos, Environ. Sci. Technol. {\bf 44}, 4539-4544 (2010).
\bibitem{17} H. Ohshima, Adv. Colloid Interface Sci. {\bf 226}, 2-16 (2015).
\bibitem{18} H. Ohshima, J. Colloid Interface Sci. {\bf 328}, 3-9 (2008).
\bibitem{24}  H. Ohshima, \emph{Biophysical Chemistry of Biointerfaces} (John Wiley $\&$ Sons, Hoboken, NJ, 2010).
\bibitem{16} T. H. Nguyen, N. Easter, L. Gutierrez, L. Huyett, E. Defnet, S. E. Mylon, J. K. Ferrid, and N. A. Viet, Soft Matter {\bf 7}, 10449 (2011).
\bibitem{19} K. McDaniel, F. Valcius, J. Andrews, and S. Das, Colloids Surf. B {\bf 127}, 143-147 (2015).
\bibitem{20} P. P. Gopmandal, S. Bhattacharyya, and H. Ohshima, Colloid. Polym. Sci. {\bf 294}, 727-733 (2016).

\bibitem{21} A. Ganjizade, S. N. Ashrafizadeha, and A. Sadeghi, Electrochem. Commun. {\bf 84}, 19-23 (2017).
\bibitem{22} S. Kumar Maurya, P. P. Gopmandal, and H. Ohshima, Colloid. Polym. Sci. {\bf 296}, 721-732 (2018).

\bibitem{23} A. Ganjizade, A. Sadeghi, and S. N. Ashrafizadeh, Colloids Surf. B {\bf 170}, 129-135 (2018).
\bibitem{5} A. Armanious, M. Aeppli, R. Jacak, D. Refardt, T. Sigstam, T. Kohn, and M. Sander, Environ. Sci. Technol. {\bf 50}, 732-743 (2016).
\bibitem{1} H. Ohshima, Curr. Opin. Colloid Interface Sci. {\bf 18}, 73 (2013).
\bibitem{2} A. Siber, A. L. Bozic, and R. Podgornik, Phys. Chem. Chem. Phys. {\bf 14}, 3746 (2012).
\bibitem{4} A. Siber and R. Podgornik, Phys. Rev. E {\bf 76}, 061906 (2007).
\bibitem{25} R. J. Nap, A. L. Bozic, I. Szleifer, and R. Podgornik, Biophys. J. {\bf 107}, 1970-1979 (2014).

\bibitem{3} E. J. Sambriski, D. C. Schwartz, and J. J. de Pablo, Biophys. J. {\bf 86}, 1675 (2009).

\bibitem{26} J.-P. Hansen and I.R. McDonald, \emph{Theory of Simple Liquids} (2nd Edition, Academic Press, London, 1986).


\end{thebibliography}
\end{document}